\documentclass[10pt,prb,aps,amssymb,amsmath,tightenlines,twocolumn,superscriptaddress]{revtex4-2} 

\usepackage{graphicx,epsfig,xcolor}
\usepackage[T1]{fontenc}
\usepackage[latin1]{inputenc}
\usepackage[english]{babel}
\usepackage{amsmath, bbm, nccmath, MnSymbol}
\usepackage{blkarray}
\usepackage{mathtools}
\usepackage[normalem]{ulem}
\usepackage{amsfonts}
\usepackage{amssymb}
\usepackage{bm}

\makeatletter
\def\maketitle{
\@author@finish
\title@column\titleblock@produce
\suppressfloats[t]}
\makeatother

\usepackage{braket}
\usepackage{slashed}
\usepackage{dsfont}
\usepackage{float}
\usepackage{color}
\definecolor{ikb}{rgb}{0, 0.184,0.655}

\usepackage[labelformat=parens]{subfig}
\usepackage[justification=raggedright, font=footnotesize]{caption}
\usepackage[colorlinks=true, linkcolor=ikb, urlcolor=ikb, citecolor=ikb,
            anchorcolor=ikb, bookmarksopen=true]{hyperref}

\usepackage{titlesec}
\titleformat{name=\section, numberless}[runin]
  {\itshape}{}{0pt}{}
  \titlespacing{\section}{0pt}{0pt}{0pt}

\newcommand{\cH}{\ensuremath{\mathcal{H}}}
\newcommand{\cP}{\ensuremath{\mathcal{P}}}
\newcommand{\cT}{\ensuremath{\mathcal{T}}}
\newcommand{\cPT}{\ensuremath{\mathcal{PT}}}

\begin{document}

\title{Parity-induced generalized Brillouin zone\\ without non-Hermitian skin effect}

\author{Alexander Felski}
\email{alexander.felski.d8@tohoku.ac.jp}
\affiliation{
Max Planck Institute for the Science of Light, Erlangen, Germany
}
\affiliation{
Advanced Institute for Materials Research, Tohoku University, Sendai, Japan}

\begin{abstract}
    Acute spectral sensitivity to boundary conditions and the formation of a generalized Brillouin zone associated with complex quasimomenta are features frequently attributed to systems with non-trivial non-Hermitian topology, showcasing the non-Hermitian skin effect.
    We show that, away from the thermodynamic limit, these features themselves are not uniquely tied to this phenomenon; they can similarly arise as parity-induced even-odd effects in non-Hermitian systems without skin effect. Despite an underlying generalized Brillouin zone description, wavefunctions remain delocalized.  
    In addition, the effect can arise in skin-effect models as entirely separate distinguishable feature.
\end{abstract}
\maketitle

\pdfbookmark[1]{Introduction}{Introduction}

\emph{Introduction.} The non-Hermitian skin effect, describing a macroscopic accumulation of states at the boundaries or interfaces of non-Hermitian systems which incorporate gain and loss or non-reciprocal coupling terms, 
is an emblematic feature in the extension of topological phases and phase transitions beyond the Hermitian framework.
The central mechanisms behind its formation are now well understood in the language of the generalized Brillouin zone and non-Bloch band theory \cite{alase2017generalization, yao2018edge, yokomizo2019non}, and its origin in the spectral topology of non-Hermitian systems \cite{okuma2020topological, gong2018topological} is well-established. Extensive symmetry categorizations based on the ramified Altland-Zirnbauer classes ($AZ$ and $AZ^\dagger$) are available \cite{kawabata2019symmetry}.
A particularly notable consequence of this localization of bulk wave functions is the acute sensitivity of the spectra of these non-Hermitian systems on the boundary conditions.

Yet, not all non-Hermitian systems display the non-Hermitian skin effect. This is prominently reflected in early studies of the paradigmatic Su-Schrieffer-Heeger (SSH) model, extended through the inclusion of alternating on-site gain and loss terms motivated by the parity-time-reflection ($\cPT$) symmetry thus realized \cite{esaki2011edge, schomerus2013topologically, zhu2014pt}. 
Despite the non-Hermiticity, the bulk wave functions of this model remain fully delocalized as in Hermitian systems. Only in later studies of non-reciprocal chains did the skin effect phenomenology come to light 
\footnote{The effect itself was already described in the earlier work of Hatano and Nelson, though its mechanism was first explored later by \cite{yao2018edge}.}.

The $\cPT$-symmetric, or more generally pseudo-Hermitian, models in fact mark a significant addition to the $AZ$ and $AZ^\dagger$ symmetry classes \cite{kawabata2019symmetry}. 
It is now widely-known that such models commonly admit regimes of real eigenvalues, wherein they behave like Hermitian models, technically captured through the introduction of a non-trivial Hilbert space metric in the biorthogonal formalism \cite{bender1998real, brody2014biorthogonal}. 
Maybe unsurprisingly then, solid-state models
in this unbroken $\cPT$ symmetry regime are also guaranteed to admit real quasimomenta $k$. That is to say, the generalized and conventional Brillouin zones coincide and, consequently, wave functions remain delocalized.
This behavior was shown to extend further to one-dimensional non-Hermitian Hamiltonians having the time-reversal symmetry (orthogonal-class variant)
\footnote{
While the conventional non-Hermitian skin effect is excluded, a subtle point of systems with $TRS^\dagger_+$ symmetry ($\text{AI}^\dagger$ class) is the possibility of a \emph{bidirectional} skin effect \cite{wang2024constraints}. This possibility is taken into account for the model system discussed.
}
\begin{equation}
    \text{TRS}^\dagger_+:\quad  \mathcal{C}_+ \cH^T(\mathbf{k}) \mathcal{C}_+^{-1} = \cH(-\mathbf{k})\, ,
    \quad \mathcal{C}_+ \mathcal{C}_+^* = +1
\end{equation}
or parity reflection symmetry
\begin{equation}
    \cP:\quad  \cP \cH(\mathbf{k}) \cP^{-1} = \cH(-\mathbf{k})\, , 
    \quad \cP^2 = 1
    \, ,
\end{equation}
even in the presence of complex eigenvalues \cite{kawabata2019symmetry, kawabata2020nonbloch}.

Here, we demonstrate the comparable formation of complex spectra with delocalized wavefunctions, while simultaneously realizing a non-trivial generalized Brillouin zone that lies in the complex quasimomentum plane and thus differs from the conventional Brillouin zone, resembling the case of the non-Hermitian skin effect. The apparent conflict is examined and resolved,
highlighting the subtle connection between skin effect and generalized Brillouin zone.  

This behavior is realized in a bipartite tight-binding chain, in which the states arising on each sublattice combine to delocalized full wave functions even in the presence of complex quasimomenta.
That is, despite displaying delocalized solutions, the underlying complex quasimomenta are essential to capturing the behavior of the system.
Moreover, and complementary to studies in the literature, this phenomenology is finite-size induced and arises as a parity-dependent even-odd phenomenon, understood to originate in the symmetry breaking of the model attainable at finite lengths, as illustrated in various Hermitian models \cite{lang1998oscillatory, yamaguchi1997even, sim2001even, machens2013even, kim2010length}. 
We highlight, that the distinct behaviors at even or odd chain lengths in the non-Hermitian model discussed in the following are not of the form of individual localized edge modes forming due to the presence of half unit cells, but due to a fundamental change of the symmetry class of the open chain.

\pdfbookmark[1]{Preliminaries}{Preliminaries}

\emph{Preliminaries.} To illustrate the generalized Brillouin zone approach and bring to mind the differences between 
boundary and bulk behavior
within this framework, we briefly recall its application to 
\begin{equation}
\label{eq:H HN}
    \cH_\text{HN} = -\sum_{n=1}^{N-1} [\,(t-g)c_{n}^\dagger c_{n+1} + (t+g)\,c_{n+1}^\dagger c_{n})\,] \, 
\end{equation}
for $g=0$ (the simple tight-binding chain $\cH_\text{tb}$) and for $g\neq0$ (the non-reciprocal tight-binding chain, or Hatano-Nelson model) \cite{hatano1996localization, nelson1993boson, hatano1997vortex}.

Following \cite{alase2017generalization}, the identification of bulk states, which coincide with the behavior obtained under assumed periodic boundary conditions (PBC), in the open boundary chain proceeds as a standing-wave construction based on the superposition of the counterpropagating states 
$
    \ket{k} = N^{-\frac{1}{2}}\sum_{n=1}^N \mathrm{e}^{i k n}\, \ket{n}
$
and $\ket{-k}$, with quasimomenta $k \in [-\pi, \pi) \subset \mathbbm{R}$.
For $\cH_\text{tb}$ these states satisfy the relations
\begin{equation}
\label{eq:tb kstates}
    \cH_\text{tb} \ket{\pm k} = -2t \cos k \ket{\pm k} + N^{-\frac{1}{2}} [ \,t \ket{1} + t \mathrm{e}^{\pm ik(N+1)} \ket{N}]\, .
\end{equation}
Boundary effects can now be cancelled to find the eigenstates $\ket{\psi_k}= \alpha\ket{k} +\beta \ket{-k}$ when $\beta = -\alpha$ (cancellation of $\ket{1}$ terms) and $k = \frac{\pi j}{N+1}$, 
$j \in [1,N] \subset \mathbbm{N}$  (cancellation of $\ket{N}$ terms), which by construction sample the PBC spectrum.
The behavior of the open chain is in full agreement with the PBC description.

\begin{figure}[t]
\centering
\includegraphics[width=0.9\columnwidth]
{
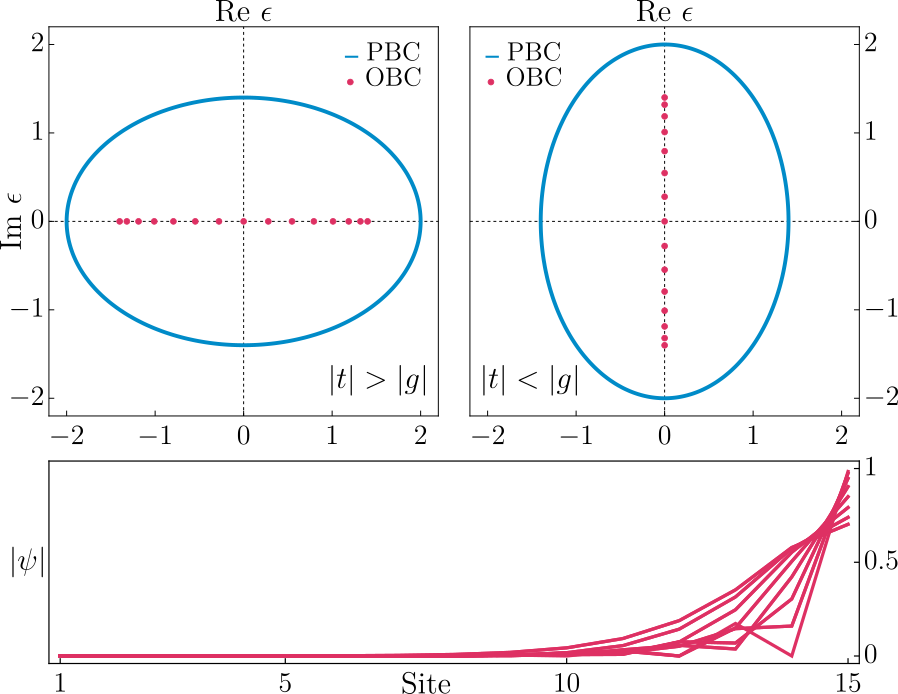
}
\caption{
\label{f1}
Features of $\cH_\text{HN}$. Top: Eigenvalue spectra under periodic (blue) and open (red) boundary conditions in the complex energy plane.   
Bottom: Macroscopic localization of bulk states under open boundaries, displaying the non-Hermitian skin effect. 
}
\end{figure}

In stark contrast to this, the PBC spectrum of the Hatano-Nelson model (\ref{eq:H HN}) obtained from its Bloch Hamiltonian formulation,
$\cH_\text{HN}(k) = -2t \cos k +2ig \sin k$,
forming an ellipse in the complex energy plane, differs dramatically from the entirely real or imaginary eigenvalues obtained in any finite OBC configuration, compare Fig.~\ref{f1}.
This is reflected in the relations
\begin{align}
\label{eq:HN kstsates}
    \cH_\text{HN} \ket{\pm k} =  \,\,&(t+g) N^{-\frac{1}{2}}\,\ket{1}
    + (t-g) N^{-\frac{1}{2}} \mathrm{e}^{\pm ik(N+1)}\,\ket{N}  \notag\\
    +&(-2 t \cos k  \pm 2ig \sin k ) \, \ket{\pm k} 
    ;
\end{align}
under the conventional assumption of plane wave bulk states, with $k \in \mathbbm{R}$, 
an approximation of the behavior by means of the Bloch Hamiltonian fails. 

This is remedied by considering an analytic deformation of the conventional Brillouin zone to an appropriate contour in the complex quasimomentum plane: 
$k\in \mathbbm{C}$, with 
$\text{Re}\,k\in [-\pi,\pi)$;
the generalized Brillouin zone \cite{yao2018edge, yokomizo2019non, zhang2020correspondence}.
Accordingly, the desired eigenstates now have the general form  
$\ket{\psi_k}= \alpha_1\ket{k} +\alpha_2 \ket{-k}+ \alpha_3\ket{k^*} +\alpha_4 \ket{-k^*}$.
For the Hatano-Nelson model
one finds that the inclusion of a constant imaginary contribution to the quasimomentum facilitates both the cancellation of boundary terms and for the bulk states to sample the PBC description
for the solutions
\begin{equation}
    \ket{\psi_k} \propto \sum_{n=1}^N N^{-\frac{1}{2}}\,r^n \, \sin[n\,\text{Re}\, k ] \, \ket{n}\, .
\end{equation}
Here $r = \sqrt{(t+g)/(t-g)} \neq 1$ is the radius of the generalized Brillouin zone. 
Notably, the remaining bulk contribution of $\cH_\text{HN}$ acting on these states is parametrized by $\text{Re}\, k$
and reduces to the form
\begin{align}
\begin{split}
\label{eq:H HN GBZ Bloch}
    \cH_\text{HN}^\text{GBZ}[\text{Re}\,k] 
    &= -2\sqrt{t^2-g^2} \cos(\text{Re}\,k) \, ,
\end{split}
\end{align}
describing the so-called non-Bloch band structure, in exact agreement with the Bloch Hamiltonian under the assumption of generalized Bloch eigenstates with radius $r$, cf. \cite{yao2018edge}.
Notable for the following discussion is the observation, that the regime $\vert t\vert < \vert g\vert$ explains the behavior of an open boundary system featuring imaginary energy eigenvalues through the use of the generalized Brillouin zone.
Moreover, due to the complex quasimomentum values on the generalized Brillouin zone, the eigenstates in either spectral phase are no longer composed of plane waves, 
reflecting the localization of all states on the boundaries, shown in Fig.~\ref{f1}; they display the non-Hermitian skin effect \cite{okuma2023non, zhang2022review, gohsrich2025non}.

\pdfbookmark[1]{Model Hamiltonian}{Model Hamiltonian}

\emph{Model Hamiltonian.} Contrasting the previous examples, we now demonstrate the possibility of a complex eigenvalue spectrum for an open boundary system, explained by a generalized Brillouin zone distinct from the conventional real quasimomentum case, which nonetheless features \emph{delocalized} eigenstates. 
The model under consideration is described by the Hamiltonian 
\begin{equation}
\label{eq:H SSH*}
    \cH_\text{SSH*} = 
    \sum_{n} [-t + (-1)^n i \delta] \, (c_{n}^\dagger c_{n+1} + c_{n+1}^\dagger c_{n}) \, ,
\end{equation}
being a non-Hermitian tight-binding model with patterned phase shift, realized through alternatingly complex conjugate coupling strengths.
Structurally, it resembles the seminal SSH model \cite{su1979solitons}, having alternating strong and weak coupling $(t+g)$ and $(t-g)$, but at imaginary dimerization strength $g\to i\delta$ (hence labeled $\text{SSH}*$).
Paralleling the SSH model, the Bloch Hamiltonian of this system is readily obtained as
\begin{equation}
\label{eq: bloch SSH*}
     \cH_\text{SSH*}(k) = \scalebox{0.85}{$
     \begin{pmatrix}
         0 & -(t\!+\!i\delta)\!-\!(t\!-\!i\delta)\mathrm{e}^{2ik} \\
         -(t\!+ \!i\delta)\!-\!(t\!-\!i\delta)\mathrm{e}^{-2ik} & 0
     \end{pmatrix}
     $}\, .
\end{equation}

Following the ramified Altland-Zirnbauer symmetry classifications \cite{kawabata2019symmetry}, this Hamiltonian has both time-reversal symmetry $\text{TRS}^\dagger$: 
$C_+ \cH_{\text{SSH}*}^T(k) C_+^{-1} = \cH_{\text{SSH}*}(-k)$, $C_+ C_+^*= {\mathbbm{1}} $
with $C_+ = \mathbbm{1}$, and TRS. For the latter one observes that 
$
\mathcal{U}\, \cH_{\text{SSH}*}^*(k)\, \mathcal{U}^{-1} = \cH_{\text{SSH}*}(-k)$, $\mathcal{U} \mathcal{U}^*= {\mathbbm{1}}
$
with $\mathcal{U} = \mathcal{U}(k) = \text{diag}(\mathrm{e}^{-ik}, \mathrm{e}^{ik})$. 
The classification requires a $k$-independent symmetry operator \cite{kawabata2019symmetry, ryu2010topological}, which is straight-forwardly identified after the (local) gauge transformation $\tilde{\cH}_{\text{SSH}*}(k) = G(k)\cH_{\text{SSH}*}(k) G^{-1}(k)$, $G(k) = \text{diag}(\mathrm{e}^{-ik/2}, \mathrm{e}^{ik/2})$, so that 
$
\mathcal{T}_+\, \tilde{\cH}_{\text{SSH}*}^*(k)\, \mathcal{T}_+^{-1} = \tilde{\cH}_{\text{SSH}*}(-k)$, $\mathcal{T}_+ \mathcal{T}_+^*= {\mathbbm{1}}$ with $\mathcal{T}_+ = \mathbbm{1}$.
Due to the topological unification of TRS and particle-hole symmetry $\text{PHS}^\dagger$, it thus falls into the class $\text{BDI}^\dagger$ and thus known to not host a non-Hermitian skin effect or any comparable topological phase.
More directly, the absence of the skin effect can be seen based on the observation that the gauge transformation $G(k)$ relates (\ref{eq: bloch SSH*}) to a Hamiltonian that is both pseudo-Hermitian and anti-pseudo-Hermitian $ \pm\eta_\pm \tilde{\cH}_{\text{SSH}*}^\dagger(k) \eta_\pm^{-1} = \tilde{\cH}_{\text{SSH}*}(k)$ with $\eta_+ = \sigma_x$ and $\eta_- = \sigma_y$. 
Thus the Bloch bands,
\begin{equation}
\label{eq: SSH* energies}
    \varepsilon(k)= \pm2\sqrt{t^2 \cos^2k-\delta^2 \sin^2k}\, ,
\end{equation}
are restricted to purely imaginary values between $\pm 2i\delta$ for quasimomenta $\pm k \in (\tan^{-1}(t/\delta), \pi-\tan^{-1}(t/\delta) )$ and purely real values between $\pm 2t$ otherwise, and consequently the model cannot host a skin effect. 

It is this $\text{BDI}^\dagger$ symmetry class, which remains preserved or breaks in odd- or even-length finite chains respectively, and thus enables distinct parity-dependent non-Hermitian phenomena.  

\begin{figure}[t]
\centering
\includegraphics[width=0.9\columnwidth]
{
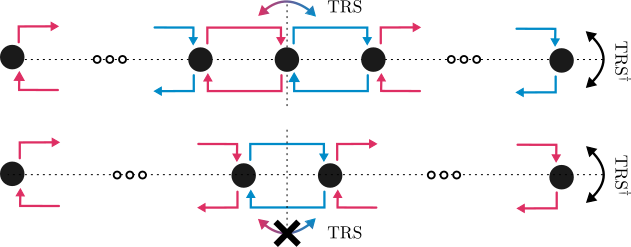
}
\caption{
\label{f2}
Schematic illustration of the $\text{TRS}^\dagger$ and $\text{TRS}$ symmetries in finite-chain realizations of $\cH_{\text{SSH}*}$ at odd (top) and even (bottom) chain length.
}
\end{figure}

\pdfbookmark[1]{Finite chains of odd length}{Finite chains of odd length}

\emph{Finite chains of odd length.} A finite-size realization of $\cH_{\text{SSH}*}$ in (\ref{eq:H SSH*}) under open boundaries and at odd length $N$ can be represented in matrix form as
\begin{equation}
    \cH_{\text{SSH}*}^\text{odd}= \scalebox{0.85}{$\begin{pmatrix}
        0 & -t_\text{eff} & 0 & & 0\\
        -t_\text{eff} & 0 &-t_\text{eff}^* & &  \\
        0 & -t_\text{eff}^* & 0 & \ddots &  \\
        & & \ddots & 0& -t_\text{eff}^*\\
        0 & & & -t_\text{eff}^* & 0
    \end{pmatrix}$}\, , \quad t_\text{eff}= t+i\delta\, .
\end{equation}
This realization is clearly invariant under transposition, identifying in position space the symmetry $\text{TRS}^\dagger$: $C_+^{-1} (\cH_{\text{SSH}*}^\text{odd})^T C_+ =\cH_{\text{SSH}*}^\text{odd}$, with $C_+ = \mathbbm{1}$. 
In addition,  $\cH_{\text{SSH}*}^\text{odd}$ is symmetric under combined complex conjugation and reflection around the central site, that is 
$\cT_+^{-1} (\cH_{\text{SSH}*}^\text{odd})^* \cT_+ = \cH_{\text{SSH}*}^\text{odd}$ with $\cT_+ = \text{antidiag(1)}$ denoting the anti-diagonal unit matrix; see also the illustration of symmetries in Fig.~\ref{f2}. As before for $\cH_{\text{SSH}*}(k)$, this identifies TRS symmetry.
In particular, in this realization of the system the boundary conditions are commensurate with a time-reversal center.
This results in the overall preserved $\text{BDI}^\dagger$ classification of odd finite chains.

\begin{figure}[t]
\centering
\includegraphics[width=0.85\columnwidth]
{
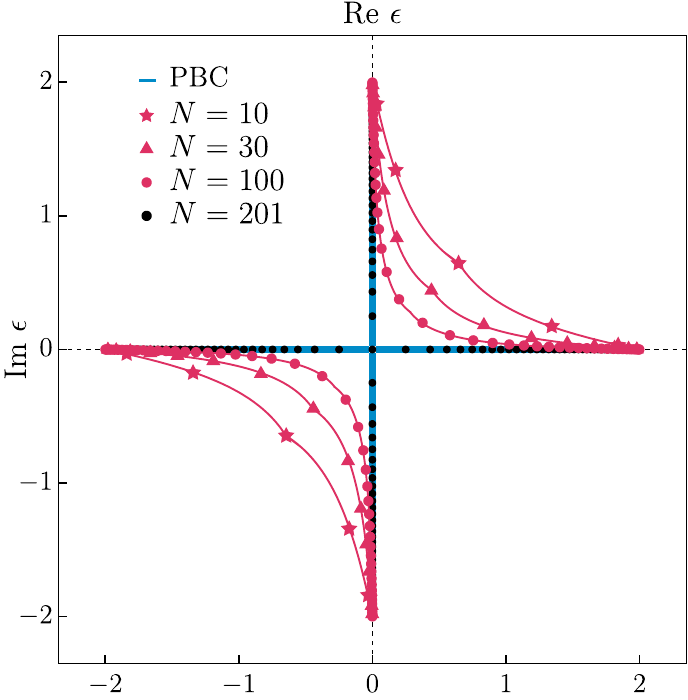
}
\caption{
\label{f3}
Spectrum of $\cH_{\text{SSH}*}$ with $t=\delta$ in the complex energy plane. Shown is the Bloch spectrum (blue), an illustrative example of the eigenvalues obtained for odd chain lengths (black), and eigenvalues found for various even chain lengths (red) displaying an acute sensitivity to the boundary conditions.
Red lines indicate the non-Bloch bands along the reconstructed generalized Brillouin zone found at even chain lengths. 
}
\end{figure}

The combination of these symmetries furthermore identifies the pseudo-Hermiticity of $\cH_{\text{SSH}*}^\text{odd}$ with $\eta_+= \text{antidiag(1)}$, and similarly $\eta_-= \text{antidiag(1,-1,1,...)}$ having alternating sign entries identifies anti-pseudo-Hermiticity.
Consequently, the eigenvalues of $\cH_{\text{SSH}*}^\text{odd}$ are again confined to the real and imaginary axes. They sample the band spectrum of $\cH_{\text{SSH}*}(k)$ and the periodic description captures the open-chain behavior well. 
Accordingly, the bulk states in the generalized Brillouin zone approach ultimately reduce to a case similar to the preliminarily discussed $\cH_\text{tb}$, yielding real-valued quasimomenta $k$, see  Supp.~Mat.~\hyperref[AppA]{A}. 
Despite the finite chain length, the bulk behavior remains unobstructed by the presence of boundaries.

\pdfbookmark[1]{Finite chains of even length}{Finite chains of even length}

\emph{Finite chains of even length.} At even length $N$ the finite-size realization of $\cH_{\text{SSH}*}$ under open boundaries breaks the TRS symmetry due to the absence of a reflection transformation around a central  
site, leaving only $\text{TRS}^\dagger$: $C_+^{-1} (\cH_{\text{SSH}*}^\text{odd})^T C_+ = \cH_{\text{SSH}*}^\text{odd}$, with $C_+ = \mathbbm{1}$, intact; the model is now 
in class $\text{AI}^\dagger$.
This class does also not host a (conventional) non-Hermitian skin effect \cite{kawabata2019symmetry, kawabata2020nonbloch}.
Nonetheless, the eigenvalues of 
\begin{equation}
\label{eq: H_SSH*_OBC}
    \cH_{\text{SSH}*}^\text{even}= \scalebox{0.85}{$\begin{pmatrix}
        0 & -t_\text{eff} & 0 & & 0\\
        -t_\text{eff} & 0 &-t_\text{eff}^* & &  \\
        0 & -t_\text{eff}^* & 0 & \ddots &  \\
        & & \ddots & 0& -t_\text{eff}\\
        0 & & & -t_\text{eff} & 0
    \end{pmatrix}$}\, , \quad t_\text{eff}= t+i\delta\, .
\end{equation}
are no longer restricted to purely real or imaginary values, but in fact move into the complex plane, cf. Fig.~\ref{f3}. 
Clearly, this behavior is no longer captured by the conventional band spectrum (\ref{eq: SSH* energies}) found under periodic boundaries. 
It marks an acute parity-induced sensitivity to the boundary conditions, reminiscent of the non-Hermitian skin effect. 

\begin{figure}[t]
\centering
\includegraphics[width=0.9\columnwidth]
{
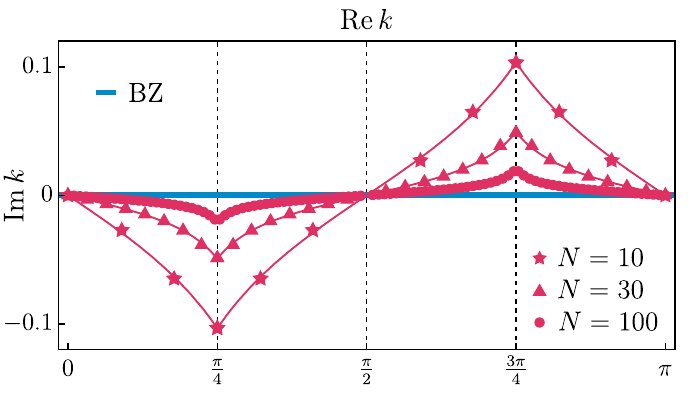
}
\caption{
\label{f4}
Complex quasimomenta of the generalized Brillouin zone description for the $\cH_{\text{SSH}*}$ model at even chain lengths and $t = \delta$. The behavior of $\text{Im}\, k$ is amplitude suppressed with increasing chain lengths, but remains structurally unchanged. The reconstructed shape of the generalized Brillouin zone is indicated as red lines. The conventional Brillouin zone is indicated in blue for reference.
}
\end{figure}

Moreover, and curiously in light of the absence of a closed-loop spectral structure under periodic boundaries, the analysis within the generalized Brillouin zone formalism identifies a consistent standing-wave construction for the description of bulk states at $k_j \in \mathbbm{C}$ with $\text{Re}\,k_j\in(0,\pi)$ satisfying
\begin{align}
\label{eq: quasimomenta_even}
\begin{split}
    \tan[k_j (N\!+\!1)] &= \mfrac{-i\delta}{t} \, \tan k_j \, , \quad
    \mfrac{\beta}{\alpha} =\mfrac{2\delta \sin k+2t \cos k}{2\delta \sin k- 2t \cos k}\, ,
\end{split}
\end{align}
see Supp. Mat. \hyperref[AppA]{A}.
The former condition yields $N$ complex-valued quasimomenta $k_j$, evaluated at which the PBC spectrum (\ref{eq: SSH* energies}) consistently reproduces the OBC energies of (\ref{eq: H_SSH*_OBC}), resembling the preliminary discussion of $\cH_\text{HN}$.
Thus, as in the non-Hermitian skin effect, the sensitivity of the spectrum to the choice of boundary conditions is captured by an underlying analytic continuation of the Brillouin zone away from the real axis:
The deviation of this generalized Brillouin zone from the conventional Brillouin zone is reflected in the imaginary part of $k$, see Fig.~\ref{f4}.
Moreover, these imaginary parts further indicate apparent exponential localization factors in the generalized Bloch waves comprising the eigenstates, that is, the \emph{apparent} presence of a finite-size induced non-Hermitian skin effect, which is not there.

We remark with regard to the finite-size nature, that this effect is:
(i) chain-length dependent and suppressed with increasing (even) $N$, in line with parity-induced even-odd effects in Hermitian models. In the thermodynamic limit $N\to \infty$, the spectrum aligns with that of the odd-length case and the periodic boundary description. A detailed discussion can be found in Supp. Mat. \hyperref[AppB]{B}.
(ii) Being chain-length dependent, one can not obtain a dense sampling of the continuous non-Bloch band structure on the generalized Brillouin zone by increasing the number of sites $N$ considered.
Nonetheless, the behavior of the generalized Brillouin zone remains structurally unchanged with increasing length and is only suppressed in the amplitude of its deviation from the conventional Brillouin zone. 
We therefore indicate its reconstructed position in addition to the set of discrete quasimomenta in Fig.~\ref{f4}, as well as the continuous non-Bloch band energies corresponding to this reconstruction in Fig.~\ref{f3}, see also Supp. Mat. \hyperref[AppB]{B}.

Despite the complex quasimomenta, the eigenfunctions are, however, not exponentially localized on the $N$ sites comprising the chain toward one boundary, like in the (conventional) non-Hermitian skin effect, but completely (skew-)symmetric \cite{cantoni1976eigenvalues}, see Supp. Mat. \hyperref[AppC]{C}.
Remarkably, the \emph{bidirectional} skin effect, in which each state is exponentially localized at both boundaries simultaneously, as discussed for a class $\text{AI}^\dagger$ system in \cite{wang2024constraints}, is in principle neither excluded by the arguments leading to the symmetry-class based categorization, nor can the (skew-) symmetry argument exclude it. 
Nevertheless, in Supp. Mat. \hyperref[AppD]{D} we demonstrate that no non-vanishing invariant like the $\text{TRS}^\dagger$ winding number of \cite{wang2024constraints} can exist for this system, indicating also the absence of bidirectional skin states, and moreover provide detailed numerical results supporting the state delocalization in the system $\cH_{\text{SSH}*}^\text{even}$.

Overall, the case of finite even-length chains demonstrates that, despite the widely-held believe of nontrivial generalized Brillouin zones and the spectral sensitivity to boundary conditions corresponding to the presence of a non-Hermitian skin effect, this is not always the case away from the thermodynamic limit $N\to \infty$.
Such effects can similarly arise as parity-induced even-odd effect in non-Hermitian systems
\footnote{We note, that this is not in disagreement with the identification of the generalized Brillouin zone as the curve of vanishing spectral winding number \cite{zhang2020correspondence}, which applies to the thermodynamic limit case only.}.

\pdfbookmark[1]{Implementation notes}{Implementation notes}

\emph{Implementation notes.} The coupling structure of the $\text{SSH}*$ model is readily implementable in current platforms for non-Hermitian systems, 
for example through the inclusion of auxiliary sites with detuning and gain or loss in a saw-tooth configuration, resulting in the desired coupling after adiabatic elimination \cite{longhi2016non}. Such configurations are commonly used to realize flux-tuning-based realizations of the Hatano-Nelson chain, e.g.~\cite{metelmann2015nonreciprocal, du2020controllable}, and are similarly able to access the $\text{SSH}*$ model phenomenology.   
In analogy to Hermitian chains that display even-odd effects in conductance oscillations depending on chain lengths \cite{lang1998oscillatory, yamaguchi1997even,sim2001even}, thus acting as parity indicator, the here discussed effect in non-Hermitian systems can potentially act as a similar signifier. 
In particular, parity changes may be determined from the contrast of the response to chiral probes at frequency $\pm \omega$. While in the spectrum of the odd chain the long-time dominating positive imaginary eigenvalues provide a symmetric contribution to both responses, the asymmetry of complex-valued spectral arcs in the even chain indicate that the long-time dominating response emphasizes the $+\omega$ probe over the $-\omega$ probe, giving rise to chiral results.
As this is determined by the imaginary eigenvalue contributions, the signal grows in amplitude over time, thus in principle providing a notable indicator even at high chain lengths.

\pagebreak

\pdfbookmark[1]{Concluding remarks}{Concluding remarks}

\emph{Concluding remarks.} We have shown that the spectral sensitivity and existence of a non-trivial Brillouin zone description of non-Hermitian models that host a non-Hermitian skin effect is not uniquely associated with such models. 
Acute spectral changes between open and periodic realizations, captured through complex quasimomentum descriptions, can arise as parity-induced effects in non-Hermitian systems away from the thermodynamic limit. Here they are not necessarily tied to the formation of localized bulk states. Despite the underlying complex quasimomenta, the resulting states can remain delocalized. 
This phenomenology is separate from the non-Hermitian skin effect, and can arise in conjunction with it (Supp. Mat. \hyperref[AppE]{E}), in which case the generalized Brillouin zone combines deviations from the conventional Brillouin zone rooted in the skin effect as well as the parity-induced features.

\pdfbookmark[1]{Acknowledgements}{Acknowledgements}

\emph{Acknowledgements.} The author is grateful to Tomoki Ozawa, Julius Gohsrich, Lukas R{\o}dland, and Flore Kunst for helpful discussions.
This work was facilitated through the Alexander von Humboldt Foundation and financially supported by JSPS Postdoctoral Fellowships for Research in Japan and JSPS KAKENHI Grant Number JP26KF0009. 
We acknowledge funding from the Max Planck Society's Lise Meitner Excellence Program 2.0.

\vfill
\bibliography{references.bib}

\clearpage

\begin{widetext}
\renewcommand{\thesection}{\Alph{section}}
\setcounter{page}{1}

\begin{center}
    \textbf{\large Supplemental Material: Parity-induced generalized Brillouin zone\\[3pt] without non-Hermitian skin effect}     
    \vspace{0.3cm}
    
    Alexander Felski,$^{1,2}$
    \vspace{0.1cm}
    
    \small
    \textit{$^1$ Max Planck Institute for the Science of Light, Erlangen, Germany} \\
    \textit{$^2$ Advanced Institute for Materials Research, Tohoku University, Sendai, Japan}
\end{center}
\vspace{0.8cm}

\section{Generalized Brillouin zone approach to the $\text{SSH}*$ model}
\label{AppA}
\vspace{0.3cm}

In applying the generalized Brillouin zone approach to the finite-length $\text{SSH}*$ model described by the Hamiltonian 
\begin{equation}
\tag{A.1}
    \cH_\text{SSH*} = 
    \sum_{n=1}^{N-1} [-t + (-1)^n i \delta] \, (c_{n}^\dagger c_{n+1} + c_{n+1}^\dagger c_{n}) \, ,
\end{equation}
one first notices that, contrary to the examples discussed in the 'Preliminaries' section, the plane wave states
$
    \ket{\pm k} = N^{-\frac{1}{2}}\sum_{n=1}^N \mathrm{e}^{\pm i k n}\, \ket{n}
$
do not satisfy a comparable eigenvalue relation with respect to $\cH_\text{SSH*}$.
Namely, acting with $\cH_\text{SSH*}$ on $\ket{\pm k }$ does not simply reproduce the scaled respective states $\ket{\pm k }$ up to some boundary terms. 
This is not surprising, since the $\text{SSH}*$ model is bipartite. 
As such, one can construct states $\ket{\pm \tilde{k}}$ that reinstate the desired behavior based on plane wave states
\begin{equation}
\tag{A.2}
    \ket{\pm k_A} = N^{-\frac{1}{2}}\sum_{\substack{n=1,\\ \text{odd}}}^N \mathrm{e}^{\pm i k n}\, \ket{n} 
    \quad \text{and} \quad
    \ket{\pm k_B} = N^{-\frac{1}{2}}\sum_{\substack{n=1,\\ \text{even}}}^N \mathrm{e}^{\pm i k n}\, \ket{n}
\end{equation}
on the respective sublattices.
These satisfy the system of relations 
\begin{align}
\tag{A.3}
\begin{split}
    \cH_\text{SSH*}\ket{\pm k_A } = \,\,& (-2t \cos k \mp 2\delta \sin k)\ket{\pm k_B}  
    + \theta(N\, \text{even})  \,\,(t-i\delta)\, N^{-\frac{1}{2}} \mathrm{e}^{\pm ik(N+1)}\,\ket{N} 
    \\[5pt]
    \cH_\text{SSH*}\ket{\pm k_B } =  \,\,& (-2t \cos k \pm 2\delta \sin k)\ket{\pm k_A} 
    + \theta(N\,\text{odd})  \,\,(t+i\delta)\, N^{-\frac{1}{2}} \mathrm{e}^{\pm ik(N+1)}\,\ket{N}  
    +(t-i\delta)\, N^{-\frac{1}{2}}\,\ket{1} \, ,
\end{split}
\end{align}
where $\theta(N\, \text{even/odd})=1$ if $N$ is even/odd and vanishes otherwise. They are reminiscent of the $\cH_\text{HN}$ relation (\ref{eq:HN kstsates}). 
Consequently, the compound states
\begin{equation}
\tag{A.4}
    \ket{\pm \tilde{k}} = \ket{\pm k_A} + \Bigl( \sqrt{\mfrac{2t \cos k + 2\delta \sin k}{2t \cos k - 2\delta \sin k}} \,\Bigr)^{\pm1}  \ket{\pm k_B}
\end{equation}
then satisfy the eigenvalue relation 
\begin{align}
\tag{A.5}
\begin{split}
    \cH_\text{SSH*}\ket{\pm \tilde{k} } =\,\,& 2\sqrt{t^2 \cos^2 k-\delta^2 \sin^2 k}\, \ket{\pm \tilde{k} } 
    +(t-i\delta)\, \Bigl( \sqrt{\mfrac{2t \cos k + 2\delta \sin k}{2t \cos k - 2\delta \sin k}} \,\Bigr)^{\pm1} N^{-\frac{1}{2}}\,\ket{1} \\
    +& \theta(N\,\text{odd})\,\,(t+i\delta)\, \Bigl( \sqrt{\mfrac{2t \cos k + 2\delta \sin k}{2t \cos k - 2\delta \sin k}} \,\Bigr)^{\pm1} N^{-\frac{1}{2}} \mathrm{e}^{\pm ik(N+1)}\,\ket{N} 
    + \theta(N\,\text{even})\,\,(t-i\delta)\, N^{-\frac{1}{2}} \mathrm{e}^{\pm ik(N+1)}\,\ket{N} \, ,
\end{split}
\end{align}
clearly being the bulk states corresponding to the PBC energies in (\ref{eq: SSH* energies}) up to boundary terms. 
The boundary terms, in contrast to the preliminary discussions, here dependent on the parity of the chain length. They can now sought to be cancelled through a superposition of counterpropagating states $\ket{\psi_k}= \alpha\ket{\tilde{k}} +\beta \ket{-\tilde{k}}$ as before. 
\\

\pdfbookmark[2]{Odd chain length N}{Odd chain length N}

\emph{Odd chain length N.} In order to cancel the boundary contributions, the coefficients $\alpha$ and $\beta$ have to be related as  
\begin{equation}
\tag{A.6}
    \beta = -\alpha\,  \mfrac{2t \cos k + 2\delta \sin k}{2t \cos k - 2\delta \sin k} 
    \quad \text{(cancellation of $\ket{1}$ terms)} \, ,
    \qquad k = \mfrac{\pi j}{N+1}, \,\,\, j \in [1,N] \subset \mathbbm{N}
    \quad \text{(cancellation of $\ket{N}$ terms)} 
    \, .
\end{equation}
This result is comparable to the preliminarily discussed tight-binding Hamiltonian $\cH_\text{tb}$. Especially noteworthy is, that the open boundary spectrum of the odd-length finite $\text{SSH}*$ chain samples the periodic description on the conventional Brillouin zone $k\in [-\pi,\pi) \subset \mathbbm{R}$. 
Accordingly, this non-Hermitian model hosts entirely delocalized wavefunctions, similar to the $\cPT$ symmetric SSH chain with alternating gain and loss, despite generally having some imaginary eigenvalues, consistent with the categorization in \cite{kawabata2019symmetry}.
\\

\pdfbookmark[2]{Even chain length N}{Even chain length N}

\emph{Even chain length N.} Parallelling the case of odd chain lengths, the cancellation of boundary terms in the even-length chain requires the relation  
\begin{equation}
\tag{A.7}
    \beta = -\alpha\,  \mfrac{2t \cos k + 2\delta \sin k}{2t \cos k - 2\delta \sin k} \, ,\quad
    \text{s.t.}\quad
    \alpha+\beta = \alpha \, \mfrac{4\delta\sin k}{-2t \cos k +2\delta \sin k} \quad  \text{and }\quad 
    \alpha-\beta = \alpha \, \mfrac{-4t\cos k}{-2t \cos k +2\delta \sin k} \, ,
\end{equation}
between $\alpha$ and $\beta$ for the cancellation of the $\ket{1}$ terms. 
However, in contrast to the previous discussion, a cancellation of the $\ket{N}$ contributions is not possible at entirely real-valued quasimomenta $k$, reminiscent of the discussion of the Hatano-Nelson model in the 'Preliminaries' section:
The condition for such a cancellation reduces to 
\begin{equation}
\tag{A.8}
    0 \overset{!}{=} \alpha (t-i\delta) \, \{
    \mfrac{4\delta\sin k}{(-2t \cos k +2\delta \sin k)} \cos[k(N+1)]
    -i \mfrac{4t\cos k}{(-2t \cos k +2\delta \sin k)} \sin[k(N+1)]
    \} \, ,
\end{equation}
that is 
\begin{equation}
\tag{A.9}
\label{eq: A9}
     \tan[k (N+1)] = -i\,  \mfrac{\delta}{t} \, \tan k \, .
\end{equation}
While this condition has no solutions for $k \in \mathbbm{R}$, it can be solved in $\mathbbm{C}$ giving rise to $N$ solutions with $\text{Re}(k) \in (0, \pi)$, thus sampling a nontrivial generalized Brillouin zone and capturing the discrepancy between the spectra obtained under open and periodic boundaries.
\\

\pdfbookmark[2]{Eigenstates}{Eigenstates}

\emph{Eigenstates.} The eigenstates of either solution thus take the general form
\begin{equation}
\tag{A.10}
\label{A.10}
    \ket{\psi_k} \propto 
    \sum_{\substack{n=1,\\ \text{even}}}^N 
    2i\, \sin(k n)
    \sqrt{\mfrac{2t \cos k + 2\delta \sin k}{2t \cos k - 2\delta \sin k}}\,
    \ket{n}
    +
    \sum_{\substack{n=1,\\ \text{odd}}}^N 
    2i\, \sin(k n)
    \ket{n}
    +
    \sum_{\substack{n=1,\\ \text{odd}}}^N 
    \mfrac{4 \delta\sin k}{-2t \cos k +2\delta \sin k}\, \mathrm{e}^{-ikn}
    \ket{n}
\end{equation}
up to an overall normalization constant.
Evidently, these functions are fully oscillatory at real-valued quasimomenta $k$, as found in the odd-length chain. But for the complex $k$ values found in the even-length case, they contain exponentially localizing factors, as is easily seen for example in the behavior on even sites $n$, scaling as $\mathrm{e}^{\vert\text{Im}k\vert n}$.
Nevertheless, the states $\ket{\psi_k}$ are not exponentially localized on the $N$ sites of the chain. A technical argument is given in Supp. Mats. \hyperref[AppC]{C} and \hyperref[AppD]{D}.
\\

\section{Thermodynamic limit and reconstructed GBZ}
\label{AppB}
\vspace{0.3cm}

\begin{figure}[b]
\centering
\subfloat[\centering $\vert t \vert \neq \vert \delta \vert$]{
\includegraphics[width=0.47\columnwidth]
{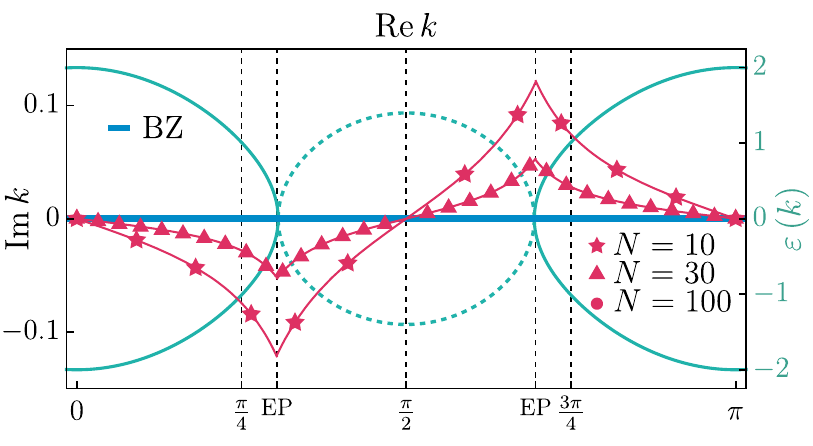}
\label{f7}
}\hspace{0.12cm}
\subfloat[\centering conjunction with skin effect]{
\includegraphics[width=0.47\columnwidth]
{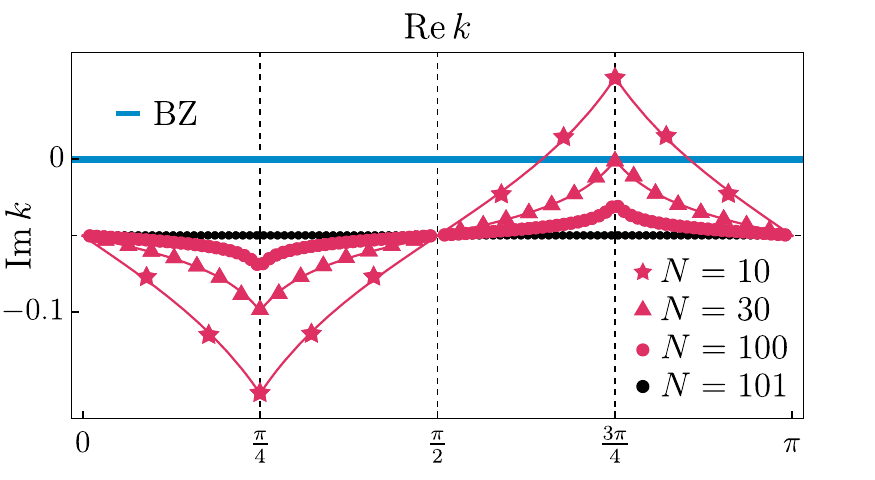}
\label{f7b}
}\hfill
\caption{
Complex quasimomenta of the generalized Brillouin zone description for the $\cH_{\text{SSH}*}$ model at even chain lengths with (a) $t=1, \, \delta=0.7$, (b) $t=\delta=1$ and inclusion of directional coupling $g=0.1$.
In (a), the real and imaginary parts of the Bloch bands (\ref{eq: SSH* energies}) are indicated as solid and dashed green lines, respectively.
}
\end{figure}

The complex-valued quasimomenta, capturing the considerable spectral deviation of even chain-length realizations of $\cH_{\text{SSH}*}$ from the results obtained under periodic boundaries or at odd lengths, are determined by the transcendental condition (\ref{eq: quasimomenta_even}), see also Eq.~(\ref{eq: A9}). 

For $t  = \delta$ the resulting behavior is shown in Fig.~\ref{f4}, displaying negative and positive imaginary quasimomentum contributions centered around $\text{Re}\, k = \frac{\pi}{4}$ and $\frac{3\pi}{4}$, respectively. (For $t=-\delta$ this behavior is simply mirrored along the real $k$ axis.) Notably, this structure remains unchanged at varying (even) chain lengths, with deviations arising only as chain-length suppressed amplitude of the imaginary quasimomentum contributions. The behavior of the imaginary part $\text{Im}\, k$, and thus the shape of the underlying generalized Brillouin zone, is generally well captured by tangent functions centered around $\text{Re}\, k = \phi \in\{0,\,\frac{\pi}{2},\,\pi \}$, i.e. $\text{Im}\,k \approx a \tan[\,b\, (\,\text{Re}\, k -\phi)\,]$. 
In Fig.~\ref{f4}, this is indicated as red lines. The non-Bloch energies along this reconstructed generalized Brillouin zone further align well with the spectral behavior, as shown in Fig.~\ref{f3}.  

For $\vert t \vert \neq \vert \delta \vert$,  the complex quasimomenta continue to follow this general behavior. However, here the regions of monotonically decreasing or increasing behavior begin to deviate from the intervals $(0,\frac{\pi}{4})$, $(\frac{\pi}{4}, \frac{3\pi}{4})$, and $(\frac{3\pi}{4},\pi)$.  
One observes that the bounds of these regions align with the changed positions of the exceptional points in the Bloch bands (\ref{eq: SSH* energies}) at $\text{Re} \, k_\text{EP} = \tan^{-1}(\pm \frac{t}{\delta}) \, \text{mod}\, \pi$. 
Figure~\ref{f7} shows this changed behavior of the complex quasimomenta, together with the Bloch band structure, for the illustrative example having $\delta=0.7t$.  

When considering the conjunction with the non-Hermitian skin effect, as modelled by 
\begin{equation}
\tag{B.1}
\label{B.1}
    \cH = \cH_\text{SSH*} + g \sum_{n=1}^{N-1} (c_{n}^\dagger c_{n+1} - c_{n+1}^\dagger c_{n}) \, ,
\end{equation}
the resulting imaginary contribution of the complex quasimomenta simply experiences an overall offset, see Fig.~\ref{f7b}. This constant imaginary contribution mirrors the behavior of the Hatano-Nelson model. The behavior of the complex quasimomenta, and thus the deviation of the generalized Brillouin zone from the conventional Brillouin zone, can be separated into contributions of the parity-induced effect, as seen in the difference of even-length chain results to the odd-length chain behavior which aligns with the thermodynamic limit, and the non-Hermitian skin effect, seen in the deviation of the odd-length chain results from the conventional Brillouin zone.
\\

Moreover, to investigate the change in the amplitude of $\text{Im}\, k$, aligning with the maximal deviation from the conventional Brillouin zone, we study the behavior of the largest and smallest imaginary contribution at a given chain length as a function of even chain lengths. The resulting behavior is shown in Fig.~\ref{f8} for the illustrative cases $\delta= t$, $\delta=0.2 t$, and $\delta = 3 t$ for chain lengths up to $N=100$.
One observes a characteristic approximate $\propto 1/N$ decay of the deviation of the generalized Brillouin zone towards the standard Brillouin zone found in the thermodynamic limit $N\to \infty$.
This qualitatively matches the behavior of parity-induced effects in Hermitian systems, even though the spectral deviation found in this non-Hermitian system is a uniquely non-Hermitian phenomenon.  

In the case of the conjunction with the non-Hermitian skin effect, the thermodynamic limit displays the same decay of the largest and smallest imaginary part of the quasimomentum. However, here the limit is determined by the non-trivial generalized Brillouin zone due to the skin effect instead of the conventional Brillouin zone in the case of $\cH_{\text{SSH}*}$, as shown in Fig.~\ref{f8b} for different strengths $g$ of the Hatano-Nelson-type contribution in (\ref{B.1}).
\\

\begin{figure}[t]
\centering
\subfloat[\centering $\vert t = \vert \delta \vert$  and $\vert t \neq \vert \delta \vert$]{
\includegraphics[width=0.47\columnwidth]
{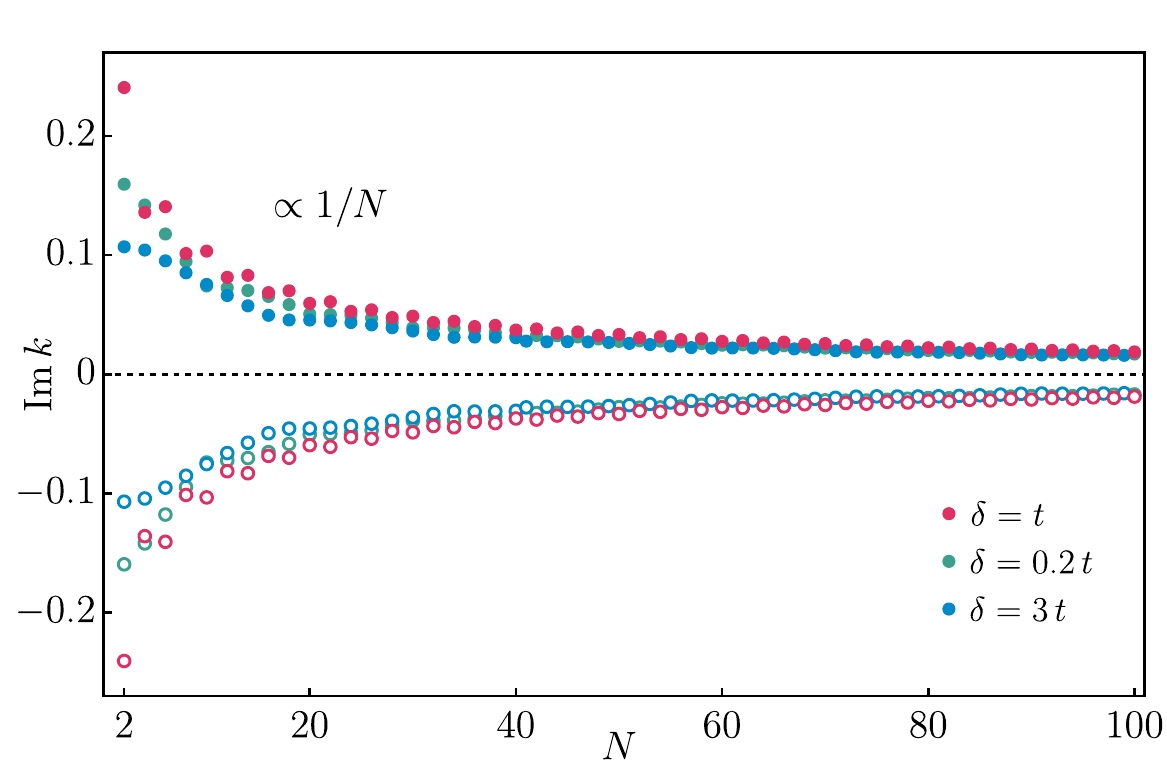}
\label{f8}
}\hspace{0.12cm}
\subfloat[\centering conjunction with the skin effect]{
\includegraphics[width=0.47\columnwidth]
{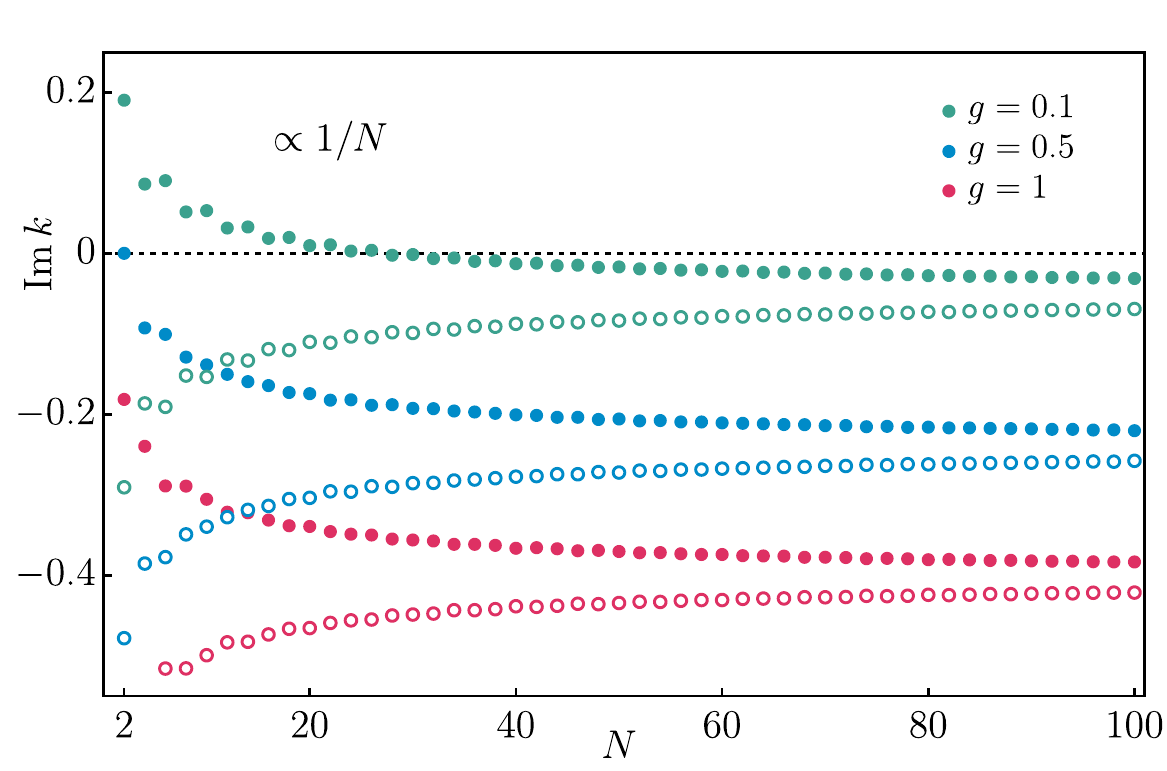}
\label{f8b}
}\hfill
\caption{
Behavior of the largest (shown as points) and smallest (shown as circles) imaginary contribution to the generalized Brillouin zone $k$ values at varying even chain lengths.
}
\end{figure}

\section{(Skew-)symmetry of eigenfunctions at even chain length}
\label{AppC}
\vspace{0.3cm}

The eigenfunctions of the even-length $\text{SSH}*$ model are completely (skew-)symmetric.
That is to say, under relabelling sites $n\to N+1-n$, the eigenfunctions remain the same (up to a possible factor of $-1$, which does not impact the localization).
This behavior can be seen based on the observation that, while being non-Hermitian, the Hamiltonian $\cH_{\text{SSH}*}^\text{even}$ in (\ref{eq: H_SSH*_OBC})
is symmetric [$\cH_{\text{SSH}*}^\text{even}$ = $(\cH_{\text{SSH}*}^\text{even})^T$], as well as persymmetric [$\cH_{\text{SSH}*}^\text{even}$ = $(\cH_{\text{SSH}*}^\text{even})^F$]. Here $(\cdot)^F$ denotes the flip-transpose $A^F = J A J $ with $J = \text{antidiag}(1)$.
Being both symmetric and persymmetric, $\cH_{\text{SSH}*}^\text{even}$ is a bisymmetric $N\times N$ matrix, and thus has $\lceil N/2\rceil$ symmetric eigenvectors ($J\psi=\psi$) and $\lfloor N/2\rfloor$ skew-symmetric eigenvectors ($J\psi=-\psi$) \cite{cantoni1976eigenvalues}.
\\

\section{Localization of the eigenfunctions at even chain length}
\label{AppD}
\vspace{0.3cm}

\begin{figure}[b]
\centering
\subfloat[\centering inverse participation ratio]{
\includegraphics[width=0.48\columnwidth]
{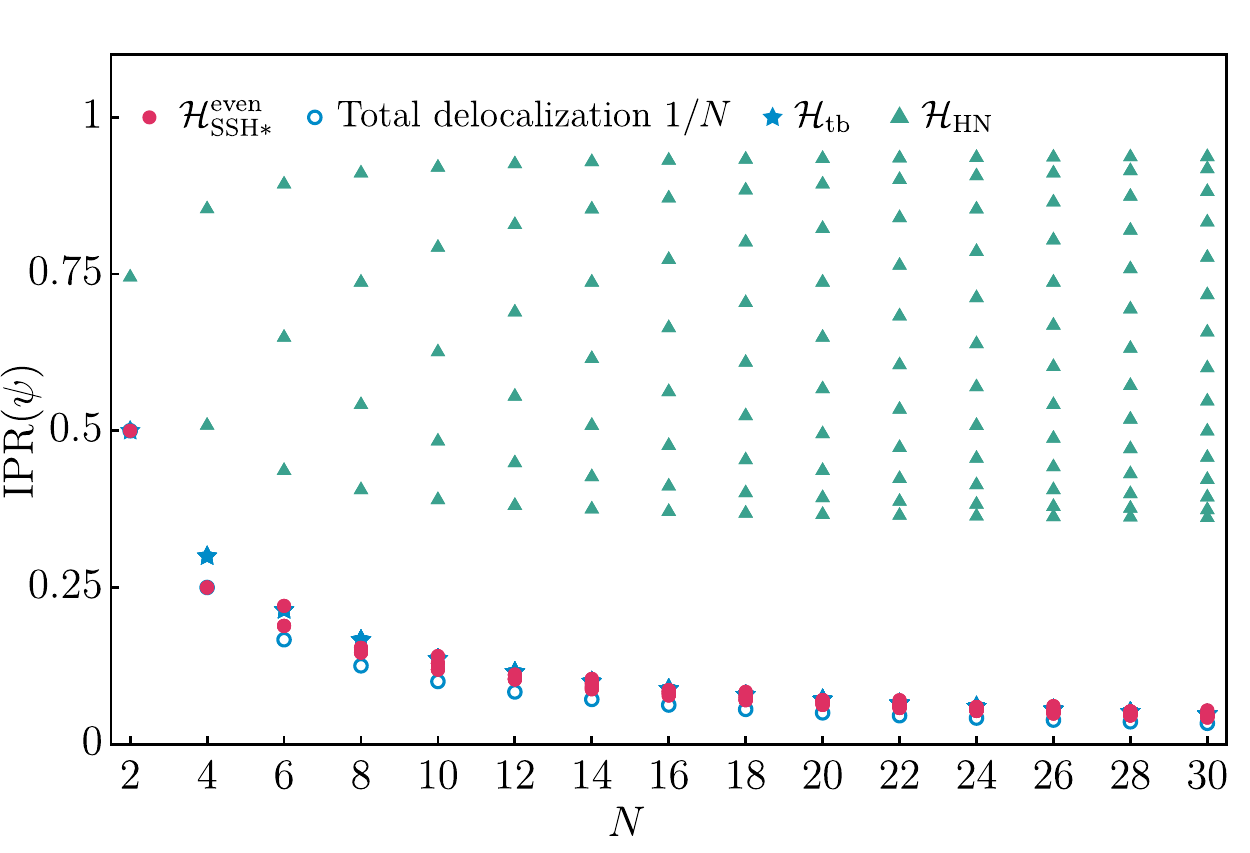}
\label{f9a}
}\hspace{0.12cm}
\subfloat[\centering participation ratio]{
\includegraphics[width=0.48\columnwidth]
{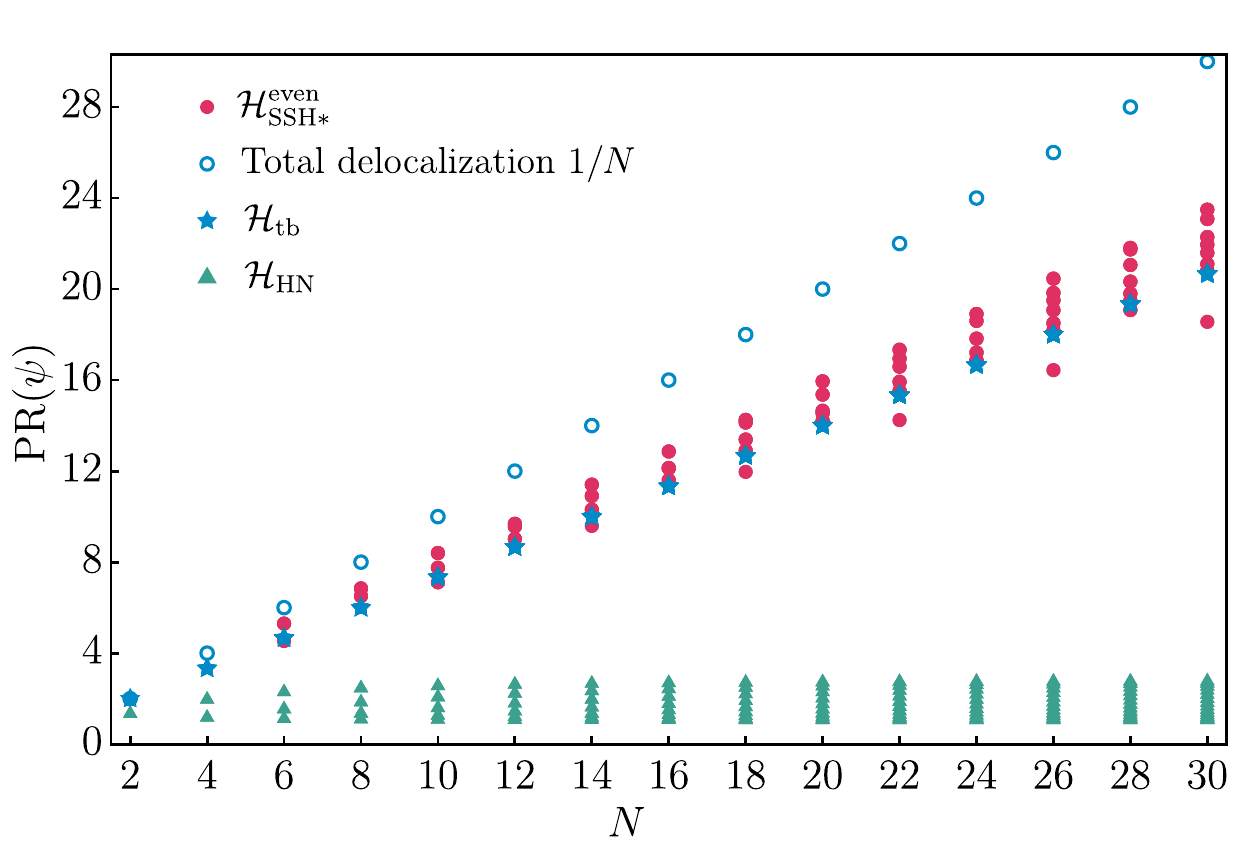}
\label{f9b}
}\hfill
\caption{
Behavior of the inverse participation ratio (a) and participation ratio (b) for the eigenstates of the $\text{SSH}*$ model at increasing even length $N$, shown as red dots. For reference the behavior of the perfectly delocalized state, the tight-binding chain eigenstates, and the exponentially localized eigenstates of the Hatano-Nelson model are visualized. 
The behavior of the even-length $\text{SSH}*$ model shows scaling typical of delocalized states.
}
\end{figure}

The (skew-)symmetry of eigenfunctions, as argued in Supp. Mat. \hyperref[AppC]{C}, does in principle not rule out the presence of a bidirectional skin effect \cite{wang2024constraints}.
This effect is associated with a non-trivial $\text{TRS}^\dagger$ winding number based on the Bloch Hamiltonian.  
While the $\text{TRS}^\dagger$ symmetry ensures that the overall winding number of system vanishes in all cases, Wang et al. showed that in the model of two coupled Hatano-Nelson chains with opposite directionality,
having the Bloch Hamiltonian 
\begin{equation}
\tag{D.1}
\label{D.1}
    \cH_\text{HN-HN}(k) = \begin{pmatrix}
        t_-\mathrm{e}^{-ik} + t_+ \mathrm{e}^{ik} & \gamma \\
        \gamma & t_-\mathrm{e}^{ik} + t_+ \mathrm{e}^{-ik} 
    \end{pmatrix} \, ,
\end{equation}
non-trivial winding numbers $\nu_\pm$ for subcontributions of the system can be introduced, which capture a bidirectional skin effect localization of eigenstates under open boundary conditions. These $\text{TRS}^\dagger$ winding numbers are exact opposites of each other, so they combine to the vanishing total spectral winding number 
\begin{equation}
\tag{D.2}
\label{D.2}
    \omega(E_b) = \mfrac{1}{2\pi i} \int_\text{BZ} \mathrm{d}k \, \partial_k \ln \det[\cH_\text{HN-HN}(k)-E_b] = 0.
\end{equation}
Nevertheless, to find possible non-trivial winding for subsets of the model still requires that the Bloch spectrum encircles the open-boundary spectrum in the complex energy plane. 
The reason for the vanishing winding $\omega$ of $\cH_\text{HN-HN}$ lies in the fact that the Bloch spectrum encircles the open-boundary spectrum in both clockwise and counterclockwise direction in the complex plane as we tune $k$ along the Brillouin zone. The non-vanishing $\text{TRS}^\dagger$ winding numbers capture exactly this structure by separating the two counteracting loops. Nevertheless, each subsystem tunes through the closed spectral loop of the Bloch Hamiltonian $\cH_\text{HN-HN}$ -- encircling the open-boundary spectrum. This latter part is an essential prerequisite to be able to find a subcontribution division that can be nontrivial. 

In the case of the $\text{SSH}*$ model, however, the Bloch spectrum is confined to the real and imaginary energy axis, see Fig.~\ref{f3}. Certainly, this structure can not encircle the complex eigenvalues found in finite even-length chain realizations. In particular, it is not possible to find any subcontribution division which will allow for non-vanishing invariants. The model does not host bidirectional skin eigenstates. 

To further substantiate the delocalization of eigenstates in the finite even-length chains $\cH_{\text{SSH}*}^\text{even}$, Figs.~\ref{f9a} and \ref{f9b} show the inverse participation ratio 
\begin{equation}
\tag{D.3}
\label{D.3}
    \text{IPR}(\psi_n) = \sum_{i=1}^N \vert \psi_{n,i} \vert^4
\end{equation}
and the participation ratio $\text{PR}(\psi_n)=\text{IPR}(\psi_n)^{-1}$ of the eigenstates of these models for increasing even chain length $N$. The behavior for a perfectly delocalized state $\text{IPR}(\psi_\text{deloc}) = 1/N$ is shown for reference. Moreover, the behavior of the tight-binding chain $\cH_\text{tb}$ is included to demonstrate the impact of finite chain lengths on the (inverse) participation ratio of delocalized states. Finally, the behavior for the non-Hermitian skin states in the Hatano-Nelson chain $\cH_\text{HN}$ is illustrated to show contrasting values for exponentially localized states. 
Notably, the eigenstates of the $\text{SSH}*$ model align well with typical behavior found in the tight-binding chain and demonstrate a scaling of the inverse participation ratio $\text{IPR}(\psi_n) \propto 1/N$ with increasing $N$, indicative of delocalized states.
\\

\section{Conjunction of the even-odd generalized Brillouin zone effect with the non-Hermitian skin effect.}
\label{AppE}
\vspace{0.3cm}

Having demonstrated the formation of a non-trivial generalized Brillouin zone that underlies the distinct behavior of finite even-length chains compared to the periodic boundary case, which correctly captured the thermodynamic limit behavior, we show here that this effect can furthermore arise additively in systems that realize the non-Hermitian skin effect. 
As preliminarily recalled, the behavior of such systems, in which the periodic description and the thermodynamic limit of the open boundary case are mismatched, is itself rooted in the generalized Brillouin zone.

When including Hatano-Nelson-type interactions (\ref{eq:H HN}) into the $\text{SSH}*$ model, 
\begin{equation}
\tag{E.1}
\label{E.1}
    \cH = \cH_\text{SSH*} + g \sum_{n=1}^{N-1} (c_{n}^\dagger c_{n+1} - c_{n+1}^\dagger c_{n}) \, ,
\end{equation}
one notes that,  on one hand, the periodic boundary spectrum, captured by the Bloch Hamiltonian 
\begin{equation}
\tag{E.2}
\label{E.2}
     \cH(k) = \cH_{\text{SSH}*}(k) + \scalebox{0.93}{$
     \begin{pmatrix}
         0 & -g (1-\mathrm{e}^{2ik}) \\
         g (1-\mathrm{e}^{-2ik}) & 0
     \end{pmatrix}
     $}\, ,
\end{equation}
prescribes a characteristic closed-loop structure in the complex energy plane.
On the other hand, the open-boundary behavior remains qualitatively unchanged from the previous discussion. In particular, the characteristic difference between odd and even chain lengths can be observed here as well, see Fig.~\ref{f5}.
The corresponding complex $k$ values sampling the generalized Brillouin zone were detailed further in Supp. Mat. \hyperref[AppB]{B}. 
For odd-length chains, one finds a constant imaginary contribution to the quasimomenta $k$, as in the Hatano-Nelson model.
But in even-length chains, the combination of this imaginary offset and the parity-induced deformation becomes apparent. 
In the thermodynamic limit, this effect is again suppressed and the asymptotic spectra of both even- and odd-length chains coincide -- but now they differ from the closed-loop spectrum of the 
periodic chain due to the presence of the non-Hermitian skin effect. Accordingly, the nontrivial generalized Brillouin zone persists in the thermodynamic limit. 
\begin{figure}[h]
\centering
\includegraphics[width=0.65\columnwidth]
{
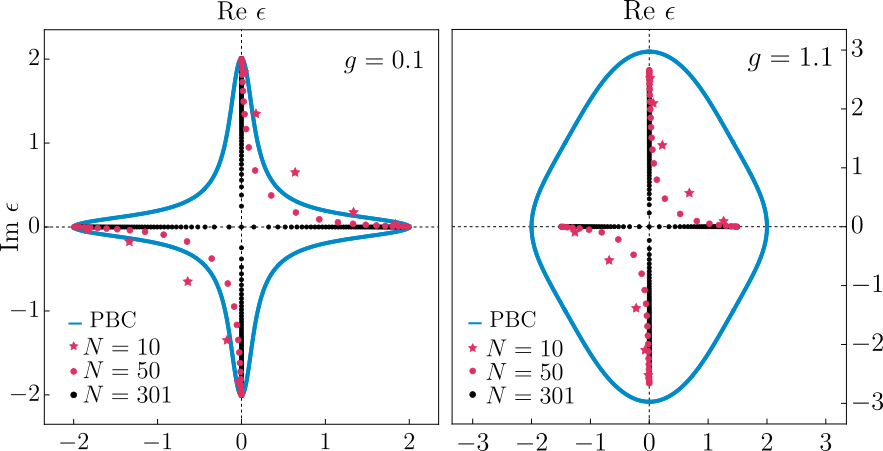
}
\caption{
\label{f5}
Spectrum of $\cH_{\text{SSH}*}$ upon inclusion of Hatano-Nelson-type directional coupling. Shown is the Bloch spectrum (blue), having the typical closed-loop structure of skin-effect models, an illustrative example of the eigenvalues obtained for odd chain lengths (black), and eigenvalues found for various even chain lengths (red), displaying an acute sensitivity to the boundary conditions.
}
\end{figure}

\clearpage\pagebreak
\vfill
\end{widetext}
\end{document}